\documentstyle[12pt,epsf]{article}
\def\beq{\begin{equation}}
\def\mpl{M_{{\rm Pl}}}
\def\eeq{\end{equation}}
\def\bea{\begin{eqnarray}}
\def\eea{\end{eqnarray}}
\def\bq{\begin{quote}}    
\def\eq{\end{quote}}

\def\bq{\begin{quote}}
\def\eq{\end{quote}}

\def\prc#1#2#3{\mbox{Phys. Rep. {\bf #1} (#2) #3}}

\begin{document}

\baselineskip 24pt

\newcommand{\sheptitle}
{Dilaton Stabilisation in $D$-term Inflation}

\newcommand{\shepauthor}
{S. F. King\footnote{On leave of absence from 
Department of Physics and Astronomy,
University of Southampton, \\ Southampton, U.K.} 
and A. Riotto\footnote{On leave of absence from 
Theoretical Physics, 
Department of Physics, University of Oxford, Oxford, U.K.}}

\newcommand{\shepaddress}
{Theory Division, CERN, CH-1211 Geneva 23, Switzerland}

\newcommand{\shepabstract}
{Dilaton stabilisation 
is usually considered to pose a serious obstacle to successful 
$D$-term inflation in  superstring theories. We argue that the physics
of gaugino condensation is likely to be modified during the inflationary phase
in such a way as to enhance the gaugino condensation scale. 
This enables dilaton
stabilisation during inflation with the $D$-term still dominating
the vacuum energy at the stable minimum. }

\begin{titlepage}
\begin{flushright}
CERN-TH/98-181\\
hep-ph/9806281\\
\end{flushright}
\vspace{.1in}
\begin{center}
{\large{\bf \sheptitle}}
\bigskip \\ \shepauthor \\ \mbox{} \\ {\it \shepaddress} \\ \vspace{.5in}
{\bf Abstract} \bigskip \end{center} \setcounter{page}{0}
\shepabstract
\begin{flushleft}
CERN-TH/98-181\\
\today
\end{flushleft}
\end{titlepage}

Due to its intrinsic elegance, inflation \cite{inflation} has become the 
almost universally accepted dogma for 
accounting for the flatness and homogeneity of the universe.
There are various classes of inflation that have been proposed,
but possibly the most successful, and certainly one of the most
popular versions these days is hybrid inflation \cite{hybrid}. 
In hybrid inflation,
there are (at least) two fields at work: the slowly rolling inflaton
field $I$, and a second field which we shall call $\psi$ whose 
value is held at zero during inflation and whose role is to end inflation
by developing a non-zero vacuum expectation value (VEV) when $I$
passes a certain critical value $I$ during its slow roll.
With $\psi=0$, the potential along the $I$ direction is approximately flat,
with the flatness lifted by a $I$ mass which must be small enough
to satisfy the slow-roll conditions for inflation. 

The natural framework for hybrid inflation is supersymmetry (SUSY).
Supersymmetry can naturally provide flat directions along which the inflaton
can roll, and additionally ensures that the scalar inflaton mass
does not have quadratic divergences. The natural size of the
inflaton mass in SUSY is of order the SUSY breaking scale, and the slow roll
conditions and COBE constraints then determine the height of the potential
during inflation $V_0^{1/4}$ to be some intermediate scale below the
grand unification (GUT) scale. To be precise,
during inflation we may set the field 
$\psi=0$ so that, assuming a flat direction, the potential simplifies to:
\begin{equation}
 V= V_0 + \frac{1}{2} m^2_I I^2.
\end{equation}
The slow roll conditions are given by:
\begin{equation}
\epsilon_N=\frac{1}{2} \mpl\left(\frac{V'}{V} \right)^2=
\frac{1}{2}\frac{\mpl^2 m_I^4 I_N^2}{V_0^2}\ll 1,
\label{epsilon}
\end{equation} 
\begin{equation}
|\eta_N | =\mpl^2 \left( \frac{V''}{V} \right) =\mpl^2
 \frac{|m_I^2|}{V_0} \ll 1.
\label{eta}
\end{equation}
The subscripts $N$ means 
that $I$ and $\epsilon$ have to be evaluated  $N\sim 50$ e-folds
before the end of inflation, when the largest scale of cosmological
interest crosses the horizon.
Here $\mpl\equiv \frac{M_{\rm P}}{\sqrt{8\pi}}=2.4\times 10^{14}$ GeV 
is the reduced Planck mass.
The spectrum of the density perturbations is given by the quantity
\cite{deltah}
\begin{equation}
\delta_H^2= \frac{32}{75}\frac{V_0}{M^4_{\rm P}}\frac{1}{\epsilon_{N}}\,,
\label{deltah}
\end{equation}
with the COBE value, $\delta_H= 1.95 \times 10^{-5}$ \cite{cobe},
expressed in terms of the (unreduced) Planck mass. 

If one accepts the above framework, one is driven almost inevitably
to supergravity (SUGRA).
The origin of the vacuum energy $V_0$ which drives inflation 
can only be properly understood within a framework which allows the
possibility for the
potential energy to settle to zero at the
global minimum, and hence lead to an acceptable cosmological constant, and
this implies SUGRA. In this way we are led to consider SUGRA models
in which the fields are displaced from their global minimum values
during inflation. 
The potential in SUGRA is given by:
\beq
V = e^G \left[ G_i (G^i_{\bar{j}})^{-1} G^{\bar{j}} - 3 \right] + V_D
\label{V}
\eeq
in the usual notation, i.e. natural units where 
$\mpl$ is set equal to unity,
with subscripts $i$ ($\bar{i}$)
referring to partial differentiation with respect to the generic 
field $\phi_i$ ($\phi_i^\ast$).
We have written the $D$-terms very schematically as $V_D$.
The K\"{a}hler function $G$ is:
\beq
G=K +\ln |W|^2, 
\label{G}
\eeq
where the K\"{a}hler potential $K$ is a real function of generic fields
$\phi , \phi^\ast$ and the superpotential $W$ is an analytic (holomorphic)
function of $\phi$ only. The K\"{a}hler metric is 
$G_i^{\bar{j}}$ and its inverse satisfies 
$(G^i_{\bar{j}})^{-1}G_k^{\bar{j}}=\delta^i_k$.
The SUGRA $F$-terms are
\beq
F^i=e^{G/2}(G^i_{\bar{j}})^{-1} G^{\bar{j}},
\label{F}
\eeq
whose non-zero value signals SUSY breaking, with a gravitino mass
\beq
m_{3/2}=e^{G/2}.
\label{m3/2}
\eeq 
Using the preceeding results we can schematically write the potential in
SUGRA as
\beq
V=|F|^2 + V_D -3m_{3/2}^2\mpl^2, 
\label{Vschematic}
\eeq
where we have put back the reduced Planck mass.
Eq. (\ref{Vschematic}) shows that there are two possible sources
for the positive 
vacuum energy $V_0$ which drives inflation: the $F$-term or the
$D$-term. The negative term also allows for eventual cancellation of the
potential energy, and is the main motivation for considering SUGRA.
If the $D$-terms are zero and the $F$-terms
of the same order of magnitude
during inflation as they are at the end of inflation
(in order to cancel the vacuum energy and lead to an acceptable
cosmological constant) we have
\begin{equation}
 V_0 = |F|^2 -3m_{3/2}^2\mpl^2
\sim m^2_{3/2} \mpl^2 \sim (10^{11}\, {\rm GeV})^4  \label{pot} \,.
\end{equation}
Here the gravitino mass is assumed to be given by,
\begin{equation}
m^2_{3/2} = < e^G > = e^K |W|^2 \sim (1 \,{\rm TeV})^2  \label{gravmass} \,,
\end{equation}
Now if we make the further assumption that
the inflaton $\phi$ mass is of order $m_{3/2}$, this line of 
reasoning leads to a violation of the slow roll condition
in Eq. (\ref{eta}); in fact from Eq. (\ref{pot})
we predict $|\eta_N | \sim 1$.
This is the so-called $\eta$ or ``slow roll'' problem.

An attractive solution to the $\eta$ problem is to suppose that
the energy which drives
inflation originates from the $D$-term \cite{D}. Then $V_D$ is allowed to take
a higher value than that in Eq. (\ref{pot}), providing it cancels
to zero at the end of inflation, thereby solving the $\eta$ problem.
In a string-inspired toy model one introduces an anomalous 
$U(1)_X$ gauge symmetry, with the fields $\phi^\pm$ having charge
$\pm 1$ while the inflation $I$ carries no anomalous charge,
and couples with a superpotential,
\begin{equation}
W=\lambda I \phi^+ \phi^-.
\end{equation}
The $D$-term in such a theory is given by
\begin{equation}
V_D=\frac{g_X^2}{2}(|\phi^+|^2-|\phi^-|^2+\xi)^2,
\end{equation}
where $\xi$ is the Fayet-Illiopoulos $D$-term related to the Green-Schwarz
mechanism of anomaly cancellation in string theories \cite{x}
\begin{equation}
\xi = \frac{g_X^2}{4}\delta_{{\rm GS}}\mpl^2
\end{equation}
and
\begin{equation}
\delta_{{\rm GS}}=\frac{1}{192\pi^2}\:{\rm Tr}\: Q_X.
\end{equation}
The idea is that during inflation the inflaton exceeds some critical value
which results in $\phi^+=\phi^-=0$ and $V_D\neq 0$,
and inflation is ended when the critical value of $I$ is reached and
allows $\phi^-$ to develop a VEV which cancels the $D$-term. Assuming that,
during inflation, the
$D$-term energy dominates over the $F$-term,
the required 
COBE normalisation may be
estimated to be \cite{review}:
\begin{equation}
\sqrt{\xi_{{\rm COBE}}}=6.6\times 10^{15}\: {\rm GeV}.
\end{equation}
From the point of view of conventional string theory this scale looks
rather low, however it may be possible to introduce additional
$U(1)$'s to ameliorate this problem \cite{string}. 
Some level of flexibility may be allowed in the case in
which, in the strong coupling limit, the ten-dimensional $E_8\otimes
E_8$ heterotic string can be described as the compactification of an
eleven-dimensional theory known as M-theory \cite{mtheory}.
When the ten-dimensional heterotic coupling is large, the fundamental
eleven-dimensional mass parameter $M_{11}$ becomes of the order of the
unification scale and it might be that the value of the 
Fayet-Illiopoulos  $D$-term may
be reduced to a value close to the one required by COBE. However, to our
knowledge no explicit example has been constructed so far.

In the context of string theories there is a further problem with 
$D$-term inflation which forms the focus of this paper, namely the
problem of dilaton stabilisation. 
In weakly-coupled string theory the gauge coupling
of $U(1)_X$ is given by
\begin{equation}
g_X^2=\frac{2}{S+S^*},
\end{equation}
where $S$ is the dilaton field whose VEV is supposed to determine the
gauge couplings. Thus in terms of the dilaton field the potential
during inflation is given by:
\begin{equation}
V_D=\frac{\delta_{{\rm GS}}^2\mpl^4}{4(S+S^*)^3}.
\label{VD}
\end{equation}
The problem is that, assuming that $V_D$ dominates over $V_F$,
the potential during inflation appears to prefer 
$s\equiv S+S^*\rightarrow \infty$ and $V_D\rightarrow 0$, 
where we have defined $s$ to be twice the real
part of $S$. This is the $D$-term inflation equivalent of the
dilaton runaway problem that appears in string theories
\footnote{Other moduli fields, like the moduli $T$, may have the same problem,
but the   $T$ run-away problem may disappear in some particular models if the 
one-loop thresholds corrections are taken into account \cite{stab}. 
In such a case the moduli $T$ is stabilized around the string scale.}.  

In the context of gaugino condensation the dilaton runaway problem may find, 
however,  a solution \cite{big4}.
In $E_8\otimes E_8$ superstring theory one of the gauge factors
is supposed to provide a hidden sector providing a natural
mechanism for breaking SUSY via gaugino condensation. 
In the effective four-dimensional theory one may have in natural units a
K\"{a}hler potential
\begin{equation}
K=-\ln(S+S^*)-3\ln(T+T^*-\phi^* \phi),
\label{K}
\end{equation}
where $T$ is a modulus field $\phi$ is some generic matter field,
and an effective  superpotential of the form
\begin{equation}
W=e^{-3S/2b_0}-c,
\label{W}
\end{equation}
where $b_0=\frac{3N_c-N_f}{16\pi^2}$ is the $\beta$-function of the 
hidden sector gauge group 
${\cal G}_{{\rm h}}$ defined by the renormalization group equation
\begin{equation}
\mu\frac{\partial g_{{\rm h}}}{\partial\mu}=-\frac{3}{2}\: b_0 \:g_{{\rm h}}^3,
\end{equation}
being $g_{{\rm h}}$ the gauge coupling constant of 
${\cal G}_{{\rm h}}$ and $\mu$ the renormalization scale.
 
The origin of this effective superpotential for the dilaton field 
is the following. In $E_8\otimes E_8$ superstring theories, 
a strongly interacting gauge theory undergoes gaugino condensation, 
\begin{equation}
\langle \bar{\chi}\chi\rangle \equiv\Lambda_{{\rm c}}^3=\mpl^3 \:e^{-3S/2b_0}.
\end{equation} 
Notice that the dynamical scale 
$\Lambda_{{\rm c}}$ carries an anomalous 
$U(1)_X$ charge since under a 
$U(1)_X$ gauge transformation with parameter 
$\alpha$ the dilaton transforms as 
$S\rightarrow S+\frac{i}{2}\delta_{{\rm GS}}\alpha$.

If one takes into account the coupling to the field strength 
$F_{\mu\nu\rho}$ of the ten-dimensional   
second-rank tensor field $a_{\mu\nu}$, 
the terms in the lagrangian that depend upon $F_{\mu\nu\rho}$ and  
${\rm Tr}[\bar{\chi}\Gamma_{\mu\nu\rho}\chi]$ form a perfect square. 
This means that the contribution of gaugino condensation to the vacuum energy 
is canceled by  $\langle F_{\mu\nu\rho}\rangle\sim
\langle F_{ijk}\rangle\sim c\:\epsilon_{ijk}$, for some complex $c$, if 
$\langle {\rm Tr}[\bar{\chi}\Gamma_{ijk}\chi]\rangle\sim 
\Lambda_{{\rm c}}^3\epsilon_{ijk}$.
The resulting  potential may be reproduced in four-dimensions by the effective
superpotential of the form (\ref{W}). 
The constant $c$, since it appears as an additive term in $W$, 
it is not renormalized. 

Now, 
the $F$-term part of the
potential which results from 
Eqs. (\ref{V}), (\ref{G}), (\ref{K}) and (\ref{W}) is
\begin{eqnarray}
V_F & = & \frac{1}{(S+S^*)(T+T^*-\phi^* \phi)^3}
\left|(S+S^*)\frac{\partial W}{\partial S}-W\right|^2 \nonumber \\
&=& \frac{1}{(S+S^*)(T+T^*-\phi^* \phi)^3}
\left|e^{-3S/2b_0}\left[1+\frac{3(S+S^*)}{2b_0}\right]-c\right|^2.
\label{VF}
\end{eqnarray}
The no-scale structure means that the  part $-3 e^G$ of the potential
has cancelled.
In the limit $c\rightarrow 0$ (or, equivalently, 
in the limit in which the gauginos do no not condense) 
the potential is minimised at zero by the
infinite runaway dilaton solution. However with a finite positive $c$
a second minimum develops also with zero energy at finite $S$,
thereby stabilising the dilaton potential in this framework.

Physically we might wish to stabilise the dilaton such that
$\alpha_{GUT}=1/24$ which would require $s=S+S^*=4$.
If the dilaton provides the dominant source of SUSY breaking
via the gravitino mass $m_{3/2}\sim 1$ TeV then we might also
wish to arrange that the gaugino condensate scale
is given by $\Lambda_{{\rm c}} = e^{-S/2b_0}= e^{-s/4b_0} \sim 10^{-6}$ 
which would imply
$b_0 \sim 0.07$, where we have assumed that the imaginary part of $S$ is
zero. One  also finds that  
$c\sim e^{-3S/2b_0}(1+\frac{3(S+S^*)}{2b_0})\sim 10^{-16}$
in natural units. Thus in the region of the minimum the typical scale
of the potential is of order $V_F \sim c^2 \sim 10^{-32}$, although
exactly at the minimum point we have $V_F=0$. By comparison the COBE
normalised $D$-term from an anomalous $U(1)_X$ as in the $D$-term inflation
scenario would be $V_D^{{\rm COBE}}=1.4\times 10^{-11}$ in natural units
(corresponding to a height of $4.6\times 10^{15}$ GeV)
which is 21 orders of magnitude larger! Thus at first sight it would appear
that, even if one accepts the above mechanism for dilaton stabilisation,
it is inadequate to resolve the dilaton problem during $D$-term inflation.

This conclusion, however,  is based on the assumption that during the
inflationary phase the
values of $b_0$ and $c$ which characterise the hidden sector are the
same as those which characterise the hidden sector in the present
non-inflationary epoch. 
The main observation of the present paper is that, clearly, they need not be, 
i.e the value of the $\beta$-function during inflation 
$b_0^{{\rm inf}}$ may be much different from its present day value 
$b_0^{{\rm tod}}$. For example  the
inflaton $I$ may couple to some hidden sector matter superfields thereby giving
them a large effective mass during inflation, and effectively decoupling them
from the $\beta$-function. By reducing $N_f$ the value of $b_0$ can be 
increased, $b_0^{{\rm inf}}\gg b_0^{{\rm tod}}$ 
and hence the gaugino condensate $\Lambda_{{\rm c}}$ and $c$ can both
be increased by orders of magnitude, thereby allowing the
dilaton to be stabilised during $D$-term inflation.
A typical example may be provided by  SUSY-QCD based on the 
gauge group $SU(N_c)$ in the hidden sector 
with $N_f\leq N_c$ flavors of ``quarks'' $Q^i$ in 
the fundamental representation and ``antiquarks'' $\widetilde{Q}_{\bar{i}}$ 
in the antifundamental representation 
of $SU(N_c)$. 
If the inflaton field $I$ gets very large values, of the order of $\mpl$,  
and it  couples to (some of)  these matter superfields in the superpotential
via $I Q^i \widetilde{Q}_{\bar{i}}$, the corresponding value of  $b_0$
will be naturally changed. This means that during inflation the values of the
gaugino condensation scale $\Lambda_{{\rm c}}$  
and the constant $c$ are different 
from the values they acquire    
in the  present true  vacuum. Indeed, during inflation,
the gauge group may undergo gaugino condensation   
at a scale larger than the traditional $10^{13}$ GeV 
and this will balance the $D$-term  driving inflation
and stabilize the dilaton $S$.

For definiteness let us consider that during inflation
the total potential is given by
\begin{equation}
V_T=V_D+V_F
\end{equation}
where $V_D$ is given in Eq. (\ref{VD}) and assume for definiteness  
that $T+T^*=1$. We assume that the $D$-term is COBE normalized, 
i.e. $V_D$ is of order $10^{-11}$ in natural units. We also assume
that $s=4$ both during inflation and after. As discussed we shall suppose that
the value of $b_0^{{\rm tod}}$ 
during inflation can be increased sufficiently that
the $F$-term dilaton stabilisation will 
stabilise the total potential $V=V_F+V_D$.
In fact the precise value of $b_0^{{\rm inf}}$ is not crucial, 
providing it is not
so small that the potential is not stabilised, or so large that $V_F$ 
dominates. Suitable typical values are $(0.2-0.25)$. For example
$b_0^{{\rm inf}}=0.2$ corresponds to a gaugino scale
$\Lambda_{{\rm c}} \sim 7\times 10^{-3}$ with the potential stabilised at $s=4$
by a constant term $c\sim 10^{-5}$. 
Since $V_F$ is 
very close to  zero at the minimum, it is obvious that the value of $V_D$ will
dominate at this point as is required in $D$-term inflation.
We find that at the minimum,
$V_D\approx 10^{-11}$ in natural units with $V_F\approx 7\times 10^{-14}$
so that $V_D$ dominates during inflation as desired.
The behavior of the three potentials $V_F,V_D,V_T$ in the inflationary phase
with the above choice of parameters is given in Fig.1.

\begin{figure}
\begin{center}
\leavevmode   
\hbox{\epsfxsize=3.5in
\epsfysize=6.6in
\epsffile{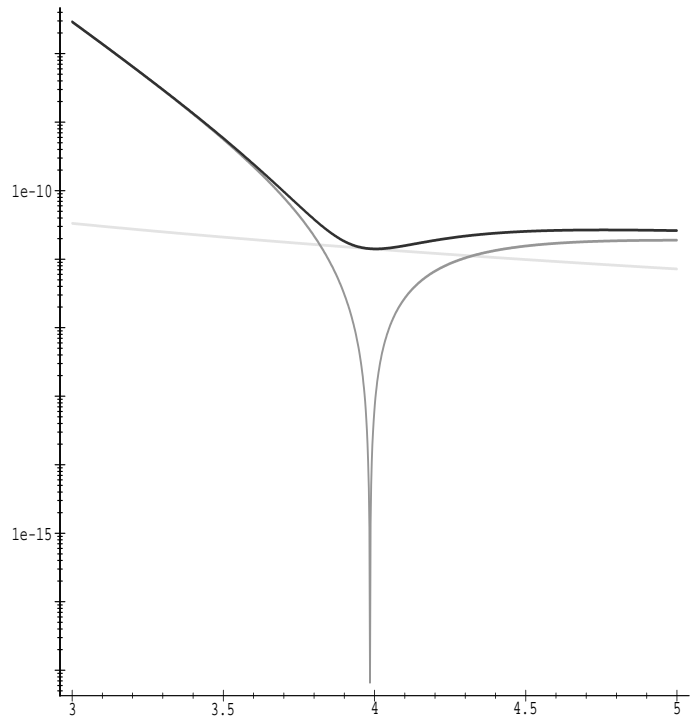}}
\end{center}
\caption{The behaviour of the potential $V_T$ (dark) 
and its component potentials $V_F$ (medium)
and $V_D$ (light) as a function of $s$
for $b_0^{{\rm inf}}=0.2$
with the potential stabilised at $s=4$
by a constant term $c\sim 10^{-5}$.  
}
\label{etas}
\end{figure}

The above values correspond to the case in which  the dilaton
is stabilised during inflation at its correct present physical value $s=4$. 
Of course, one might envisage situations 
where, during inflation, it is not sitting at its present value, 
but our assumption represents the most economical and surely 
the most harmless one. If for example 
the dilaton were stabilized at large values 
during inflation by gaugino condensation, $s\gg 1$, 
it would run away to infinity after inflation 
when $b_0$ assumes its present value $b_0^{{\rm tod}}$,  
with the dilaton separated from the true physical 
value $s=4$ at the minimum of the potential 
by an insurmountable barrier. Our assumption clearly avoids this.

One may wonder how the perfect square approach to dilaton stabilisation
survives in strongly coupled theories such as M-theory. This question
has recently been addressed by a number of authors \cite{lalak}.
The answer seems to be inconclusive. What happens is that the effective
superpotential receives additional corrections, and in particular
develops a $T$ dependence with $S$ replaced by $S+\alpha T$.
The presence of the $T$ dependence violates the no-scale structure
of the potential which is no longer a perfect square. However it is
possible that dilaton stabilisation survives even though the
potential may no longer be zero at the minimum. Another effect of
M-theory is to split the gauge couplings in the hidden sector $g_{{\rm h}}$
from the observable sector $g_{{\rm o}}$ with:
\begin{equation}
g_{{\rm o}}^2=\frac{2}{s+\alpha t}, \ \  g_{{\rm h}}^2=\frac{2}{s-\alpha t}.
\end{equation}
Thus the hidden sector gauge coupling $g_{{\rm h}}$ may be naturally larger
than in weakly coupled heterotic string theories. This feature means that
the values of both $b_0^{{\rm tod}}$ and
$b_0^{{\rm inf}}$ will be smaller in these theories than the values
we estimated earlier. In particular the value of $b_0^{{\rm inf}}\sim 0.2$
which may seem rather large may be reduced to a more comfortable 
value in M-theory.

To summarise, our
mechanism for dilaton stabilisation in $D$-term inflation is based
on a particular mechanism for $F$-term dilaton stabilisation in the
weakly coupled heterotic string, namely to appeal to hidden sector
gaugino condensation in the presence of the field strength $F_{\mu \nu \rho}$
which results in the perfect square form of the potential, and the
cancellation of the vacuum energy for a particular value of the dilaton field.
We have pointed out that this same mechanism to stabilise the
total potential can be used 
during inflation, since the scale of gaugino
condensation during the inflationary phase may be much higher than in the
physical vacuum so that $V_F$ may be increased to of order
$V_D$ in the vicinity of the minimum.
This allows usual $D$-term inflation to proceed, and answers the question
of dilaton stabilisation both during and after inflation. Finally we
point out that a similar
idea may be used for any situation in which the dilaton
is stabilised in the $F$-term of the potential. For example 
there may be non-perturbative corrections to the K\"{a}hler potential
of the dilaton which serve to stabilise the dilaton potential
(for a particular parametrisation see ref. \cite{dilaton}.)
Clearly the strength of such non-perturbative corrections may 
differ between the inflationary phase and the present vacuum, and so
one could envisage an analagous scenario to that presented here in which
the total potential is stabilised during inflation by such non-perturbative
effects.

\begin{center}
{\bf Acknowledgements}
\end{center}
We would like to thank E. Dudas and G.G. Ross
for useful conversations.

\end{document}